\documentclass{iopart}

\usepackage{amsopn,amscd,amssymb,amsthm,epsf}
\usepackage{mathptmx,courier}

\DeclareMathOperator{\End}{End}

\DeclareMathOperator{\cov}{span}

\newcommand{\setC}{\mathbb{C}}

\newtheorem{prop}{Proposition}

\newtheorem{lem}{Lemma}
\newtheorem{df}{Definition}
\theoremstyle{remark}
\newtheorem{rem}{Remark}
\newtheorem{exa}{Example}

\begin{document}
\bibliographystyle{unsrt}

\title{Destruction of states in quantum mechanics}

\author{P Caban, J Rembieli{\'n}ski, K A Smoli{\'n}ski and Z Walczak}

\address{Department of Theoretical Physics, University of {\L}{\'o}d{\'z},
  ul.~Pomorska 149/153, 90-236~{\L}{\'o}d{\'z}, Poland}


\begin{abstract}
  A description of destruction of states on the grounds of quantum
  mechanics rather than quantum field theory is proposed.  Several
  kinds of maps called supertraces are defined and used to describe
  the destruction procedure.  The introduced algorithm can be treated
  as a supplement to the von Neumann--L{\"u}ders measurement.  The
  discussed formalism may be helpful in a description of EPR type
  experiments and in quantum information theory.
\end{abstract}



\section{Introduction}
\label{sec:introduction}

In this paper we propose a solution to the following problem: how to
describe a destruction of a particle on the level of quantum mechanics
with finite degrees of freedom.  This question arises when
Einstein--Podolsky--Rosen type experiments \cite{einstein35} (see also
e.g., \cite{aspect82}) or the tests of quantum mechanical state
reduction (see, e.g., \cite{d'ariano99}) are studied.  In this type of
experiments two particles are produced in an entangled state and sent
to two measurement devices in the distance where correlated quantities
are measured at the same time.  Prediction of the correlation between
the data does not cause any problems in such an ideal experiment, but
if both measurements are not really performed at the same time we have
to take into account that a particle is irreversibly absorbed by a
detector during the measurement.  This has nothing in common with an
annihilation of a particle in quantum field theory; therefore, to
avoid any confusion we shall use the word ``destruction'' to name this
kind of processes.

Evidently, if we take into account the destruction we have to consider
open quantum mechanical systems.  We make the idealization relying on
the assumption that the destruction process is instantaneous,
therefore its description should not involve any dynamics.  For this
reason the methods of quantum field theory are not appropriate for our
purpose since QFT can be applied to open systems only if the dynamics
is given, e.g.\ by coupling the fields to external classical sources.
Moreover, in QFT formalism one has to use an infinite direct sum of
tensor product Hilbert spaces (asymptotic Fock space) while we would
like to describe quantum systems with finite degrees of freedom.

Destruction of a particle in a detector usually occurs when some
quantum numbers (e.g.\ spin, position or momentum) of the particle
belong to a specified subset of spectrum of the corresponding
observable.  Therefore, we must have a quantum system and a detector
which checks if the particle quantum numbers are inside a given subset
of spectrum.  If the answer is ``yes'', the particle is destroyed.

In this paper, we introduce a mathematical framework which allows us
to define destruction process based on the principles of quantum
mechanics.  The physical examples of destruction, including spatial
localization of particles as well as application of the destruction to
calculation of quantum correlations will be given in the forthcoming
papers.

The paper is organized as follows.  In \sref{sec:destr-one-part} we
consider destruction of one-particle state, first intuitively, then
formally.  In the next section we discuss the space of states
necessary for the description of destruction of two-particle states.
In \sref{sec:operator-traces} we introduce supertraces and study their
basic properties.  The sections \ref{sec:destr-syst-two-d} and
\ref{sec:destr-syst-two-i} deal with the destruction of two-particle
systems of distinguishable and identical particles, respectively.  We
illustrate each of these cases by examples.

\section{Destruction of one particle}
\label{sec:destr-one-part}

We begin with the discussion of a toy model in which the destruction
of a single particle takes place in a given region of space.  In the
framework of this model we formulate a description of the process of
destruction of a one-particle state taking the physical intuition as a
guiding principle.  And then we consider the general case, not
necessary related to the localization of particle.

\begin{figure}
  \begin{center}
    \epsfbox{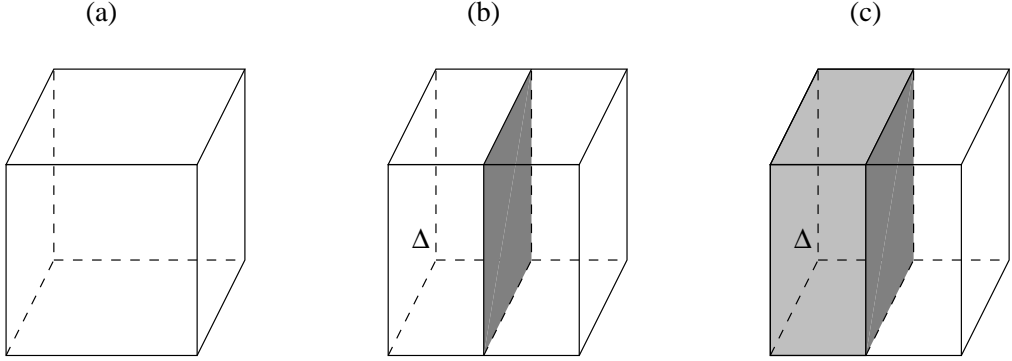}
  \end{center}
  \caption{Destruction of a particle in a part of a box: (a) there is a 
    particle in the box, (b) the box is divided by a barrier, (c)
    destruction in the region $\Delta$---there is no particle in the gray
    part of the box.}
  \label{fig:box}
\end{figure}

Thus, let us consider a box containing one particle (see
\fref{fig:box}(a)) in the state given by the density matrix $\rho$.
Now we divide the box into two parts (e.g.\ by a non-penetrating
barrier---\fref{fig:box}(b)).  We destroy the particle if it is inside
the region $\Delta$ of the box (\fref{fig:box}(c)).

First, let us discuss the situation when we check if the particle is
inside $\Delta$.  It means that we first perform a measurement with
selection of the observable $\Pi_\Delta$, where $\Pi_\Delta$ is the
projector onto the subspace of the states localized in $\Delta$.  The
measurement of $\Pi_\Delta$ gives either $1$ if the particle is inside
the region $\Delta$, or $0$ if it is outside $\Delta$.  The particle
is destroyed if the measurement of $\Pi_\Delta$ gives $1$, i.e.\ its
state is replaced by the vacuum state.  Thus, in this case, the
destruction procedure is done in two \emph{immediate} steps:
\begin{enumerate}
\item the initial density matrix $\rho$ is reduced to
  \begin{equation}
    \label{eq:20}
    \rho' = \cases{\frac{\Pi_\Delta \rho \Pi_\Delta}{\Tr(\rho \Pi_\Delta)} 
      & if the particle 
      is inside \(\Delta\)\\
      \frac{\Pi_\Delta^\perp \rho \Pi_\Delta^\perp}%
      {\Tr(\rho \Pi_\Delta^\perp)} & if the particle is outside \(\Delta\)}
  \end{equation}
  where $\Pi_\Delta^\perp = I - \Pi_\Delta$ ($I$ denotes the identity
  operator);
\item if $\rho' = \Pi_\Delta \rho \Pi_\Delta/\Tr(\rho \Pi_\Delta)$,
  then it is mapped onto vacuum density matrix $\rho_{\rm vac}$,
  otherwise it is left unchanged, so
  \begin{equation}
    \label{eq:21}
    \rho'' = \cases{\rho_{\rm vac} & particle inside \(\Delta\)\\
      \frac{\Pi_\Delta^\perp \rho \Pi_\Delta^\perp}%
      {\Tr(\rho \Pi_\Delta^\perp)} & particle outside \(\Delta\).}
  \end{equation}
\end{enumerate}

But what happens if we put the barrier, but we would have not checked
if the particle was inside $\Delta$?  This situation corresponds to a
measurement with no selection of the observable $\Pi_\Delta$.  The
particle is either inside $\Delta$ with the probability $\Tr(\rho
\Pi_\Delta)$ or outside $\Delta$ with the probability $\Tr(\rho
\Pi_\Delta^\perp)$, thus
\begin{enumerate}
\item first, the density matrix $\rho$ is reduced to
  \begin{equation}
    \label{eq:26}
    \rho' = \Pi_\Delta \rho \Pi_\Delta 
    + \Pi_\Delta^\perp \rho \Pi_\Delta^\perp,
  \end{equation}
\item then, after the destruction we get either the vacuum with the
  probability $\Tr(\rho \Pi_\Delta)$ or the one-particle state with the
  probability $\Tr(\rho \Pi_\Delta^\perp)$, so
  \begin{equation}
    \label{eq:27}
    \rho'' = \Pi_\Delta^\perp \rho \Pi_\Delta^\perp 
    + \Tr(\rho \Pi_\Delta) \rho_{\rm vac}.
  \end{equation}
\end{enumerate}

It is easy to see that in the both cases the map $\rho \mapsto \rho''$
is linear on the combinations $\mu \rho_1 + (1 - \mu) \rho_2$, where
$\mu \in [0, 1]$ and $\rho_1, \rho_2$ are the density matrices, i.e.
in the convex set of density matrices.

Now, let us rewrite the above procedure in a slightly more abstract
and general context, not necessarily related to the localization of a
particle.  Let $\mathcal{H}$ be the Hilbert space of states for a
particle.  The one-particle states (density matrices) form a convex
subset of the endomorphism space of $\mathcal{H}$ (i.e.\ $\rho \in
\End(\mathcal{H})$).  In order to describe the system if the
destruction occurs we must introduce the vacuum state $|0\rangle$ and
one-dimensional vacuum space spanned by $|0\rangle$, i.e.\ 
$\mathcal{H}^0 \equiv \cov\{|0\rangle\}$.  The vacuum vector
$|0\rangle$ is orthogonal to any vector from $\mathcal{H}$ and every
observable acts trivially on it.  Therefore, the Hilbert space of the
system under consideration is a direct sum $\mathcal{H} \oplus
\mathcal{H}^0$, and the states are mixtures of the elements from
$\End(\mathcal{H})$ and $\End(\mathcal{H}^0)$.  Furthermore, let
$\hat{\Lambda}$ be an arbitrary observable with the spectrum $\Lambda$
and $\Omega$ be a subset of the spectrum.  Denote the subspace spanned
by all the eigenvectors corresponding to the eigenvalues from the
subset $\Omega$ by $\mathcal{H}_\Omega$ and the projector onto this
subspace by $\Pi_\Omega$.  If the particle state is an element of
$\End(\mathcal{H}_\Omega)$ then the particle is destroyed, otherwise
it is not.

Therefore, let us find linear map from $\End(\mathcal{H})$ to
$\End(\mathcal{H}^0)$ which leaves the trace invariant.  It is enough
to restrict ourselves to the endomorphisms of the form $|\chi\rangle
\langle\phi|$, where $|\chi\rangle, |\phi\rangle \in \mathcal{H}$.
This map must act on these endomorphisms in the following way:
\begin{equation}
  \label{eq:3}
  \End(\mathcal{H}) \ni |\chi\rangle \langle\phi| 
  \mapsto c |0\rangle \langle0| \in \End(\mathcal{H}^0).
\end{equation}
Because $\Tr(|\chi\rangle \langle\phi|) = \langle\phi|\chi\rangle$ and
$\Tr(c |0\rangle \langle0|) = c$, it follows that $c =
\langle\phi|\chi\rangle$.  Therefore, this leads to the following
definition.
\begin{df}
  \label{df:op-trace}
  The supertrace $\widehat{\Tr}$ is a linear map
  $\widehat{\Tr}\colon \End(\mathcal{H}) \to \End(\mathcal{H}^0)$ such
  that its action on the endomorphism of the form $|\chi\rangle
  \langle\phi| \in
  \End(\mathcal{H})$ is given by the following formula
  \begin{equation}
    \label{eq:4}
    \widehat{\Tr}(|\chi\rangle \langle\phi|) 
    = \langle\phi|\chi\rangle |0\rangle \langle0|.
  \end{equation}
\end{df}
We call $\widehat{\Tr}$ supertrace\footnote[1]{We point out to avoid a
  confusion that this supertrace has nothing common with the
  supertrace ${\rm Str}$ used in supersymmetry.}  because it is a
superoperator, i.e.\ it is the operator in the endomorphism space (see
e.g.\ \cite{caves99}).

It is easy to check that if the set of vectors $\{|a\rangle\}$ is an
orthonormal basis\footnote{If we consider continuous bases, we must
  replace sums and Kronecker deltas by integrals and Dirac deltas,
  respectively.} in $\mathcal{H}$ and {$\hat{L} = \sum_{aa'} L_{aa'}
  |a\rangle \langle a'| \in \End(\mathcal{H})$} is a linear operator,
then
\begin{equation}
  \label{eq:5}
  \widehat{\Tr}(\hat{L}) 
  = \sum_{aa'} L_{aa'} \delta_{aa'} |0\rangle \langle0| 
  = \Tr(\hat{L}) |0\rangle \langle0|
\end{equation}
($\delta_{aa'}$ denotes the Kronecker delta).

Applying the $\widehat{\Tr}$ operation to the $\Omega$-projected part
of $\rho$ (i.e.\ $\Pi_\Omega \rho \Pi_\Omega$) we can formalize the
procedure which gave us the density matrix $\rho''$ by the following
definitions.

\begin{df}
  \label{df:destr-sel}
  A destruction with selection in the set $\Omega$ of one-particle
  state $\rho \in \End(\mathcal{H})$ is defined by the map\footnote{We
    shall use frequently the more general and shorter term ``map''
    instead of ``superoperator'' if it does not lead to
    misunderstandings.}  $D^s_\Omega\colon \End(\mathcal{H}) \to
  \End(\mathcal{H}) \oplus
  \End(\mathcal{H}^0)$, such that
  \begin{equation}
    \label{eq:36}
    D^s_\Omega(\rho) 
    = \cases{\frac{\widehat{\Tr}(\Pi_\Omega \rho \Pi_\Omega)}%
      {\Tr(\rho \Pi_\Omega)} 
      & if the measurement of \(\Pi_\Omega\) gives \(1\)\\
      \frac{\Pi_\Omega^\perp \rho \Pi_\Omega^\perp}%
      {\Tr(\rho \Pi_\Omega^\perp)} & if the measurement of \(\Pi_\Omega\) 
      gives \(0\).}
  \end{equation}
\end{df}

\begin{df}
  \label{df:dest1}
  The destruction with no selection in the set $\Omega$ of
  one-particle state $\rho \in \End(\mathcal{H})$ is defined by the
  map $D_\Omega\colon \End(\mathcal{H}) \to \End(\mathcal{H}) \oplus
  \End(\mathcal{H}^0)$, such that
  \begin{equation}
    \label{eq:6}
    D_\Omega(\rho) = \Pi_\Omega^\perp \rho \Pi_\Omega^\perp 
    + \widehat{\Tr}(\Pi_\Omega \rho \Pi_\Omega).
  \end{equation}
\end{df}
Note that $D^s_\Omega$ and $D_\Omega$ are superoperators.  In quantum
information theory superoperators similar to $D_\Omega$ are considered
as choice superoperators describing the coherent information transfer
between subsets of the entire system \cite{grishanin00}.

It is easy to check that applying the destruction maps $D^s_\Omega$
and $D_\Omega$ to the density matrix $\rho$ describing a state of a
particle in a box (see above), we get the density matrices $\rho''$
from \eref{eq:21} and \eref{eq:27}, respectively, when $\hat{\Lambda}$
is the position operator, $\Omega = \Delta$ and $\rho_{\rm vac} =
|0\rangle \langle0|$.

We have to show that the endomorphisms $D^s_\Omega(\rho)$ and
$D_\Omega(\rho)$, which we get after the destruction, are density
matrices.  In other words, we have to prove that $D^s_\Omega$ and
$D_\Omega$ are Kraus maps \cite{kraus83}.  This is guaranteed by the
following proposition.
\begin{prop}
  \label{prop:dest1}
  The superoperators $D^s_\Omega$ and $D_\Omega$ from the
  definitions~\textup{~\ref{df:destr-sel}}
  and~\textup{\ref{df:dest1}}, respectively, are Kraus maps.
\end{prop}
\begin{proof}
  Indeed, $D^s_\Omega(\rho)$ and $D_\Omega(\rho)$ are Hermitian
  because $\Pi_\Omega^{\dag} = \Pi_\Omega$, and $\Tr(\rho \Pi_\Omega)$
  and $\Tr(\rho \Pi_\Omega^\perp)$ are real.  Next,
  $\Tr\left(\widehat{\Tr}(\Pi_\Omega \rho \Pi_\Omega)\right) =
  \Tr(\rho \Pi_\Omega)$, so $\Tr(D^s_\Omega(\rho)) = 1$. Because
  $\Pi_\Omega^\perp = I - \Pi_\Omega$, we have
  \begin{equation}
    \label{eq:32}
    \Tr(D_\Omega(\rho)) = \Tr(\rho \Pi_\Omega^\perp) + \Tr(\rho \Pi_\Omega) 
    = \Tr(\rho) = 1.
  \end{equation}
  The proof that $D^s_\Omega(\rho)$ and $D_\Omega(\rho)$ are
  non-negative is obvious.  $\Pi_\Omega^\perp \rho \Pi_\Omega^\perp$
  is non-negative because it is an orthogonal projection of a
  nonnegative $\rho$.  $\widehat{\Tr}(\Pi_\Omega \rho \Pi_\Omega) =
  \Tr(\rho \Pi_\Omega) |0\rangle \langle0|$ is non-negative because
  $\Tr(\rho \Pi_\Omega) \geq 0$.  Thus $D^s_\Omega(\rho)$ is
  non-negative.  $D_\Omega(\rho)$ is also non-negative, because it is
  the sum of two non-negative terms, which act in orthogonal
  subspaces.  So the maps $D^s_\Omega$ and $D_\Omega$ are Kraus maps.
\end{proof}

We now illustrate the destruction procedure in the case when the
observable $\hat{\Lambda}$ is not the position operator by the
following example.
\begin{exa}
  \label{exa:destr-1}
  Consider a spin-$\frac{1}{2}$ particle.  We assume that the
  destruction \emph{with no selection} takes place if the
  $z$-component of the spin is $\frac{1}{2}$.  In this case
  $\hat{\Lambda} = \hat{S}_3$ and its spectrum is $\Lambda =
  \left\{-\frac{1}{2}, \frac{1}{2}\right\}$ and $\Omega =
  \left\{\frac{1}{2}\right\}$.  The one-particle Hilbert space is
  $\mathcal{H} = \cov\{|\uparrow\rangle, |\downarrow\rangle\}$, the
  subspace $\mathcal{H}_\Omega = \cov\{|\uparrow\rangle\}$ and the
  corresponding projection operator is $\Pi_\Omega = |\uparrow\rangle
  \langle\uparrow|$, so $\Pi_\Omega^\perp = |\downarrow\rangle
  \langle\downarrow|$.  The most general density matrix in this case
  is
  \begin{displaymath}
    \rho = w |\uparrow\rangle \langle\uparrow| 
    + c |\uparrow\rangle \langle\downarrow| 
    + c^* |\downarrow\rangle \langle\uparrow| 
    + (1 - w) |\downarrow\rangle \langle\downarrow|
  \end{displaymath}
  where $w \in [0, 1]$, $c \in \setC$ and $|c|^2 \leq w (1 - w)$.  After the
  destruction we get the new state
  \begin{displaymath}
    D_\Omega(\rho) = w |0\rangle \langle0| 
    + (1 - w) |\downarrow\rangle \langle\downarrow|.
  \end{displaymath}
  So we get vacuum state with the probability $w$ and the particle
  with $S_3 = -\frac{1}{2}$ with the probability $1 - w$.
  
  In this case it is easy to find the von Neumann entropy of the state
  before and after the destruction.  The eigenvalues of $\rho$ are
  $\rho_\pm = \frac{1}{2} \pm \sqrt{\left(\frac{1}{2} - w\right)^2 +
    |c|^2}$, so the von Neumann entropy before destruction is
  \begin{displaymath}
    S(\rho) = -\Tr(\rho \ln \rho) = -\rho_+ \ln\rho _+ - \rho_- \ln \rho_-.
  \end{displaymath}
  Because $\partial S(\rho)/\partial|c|^2 \leq 0$ for $0 \leq |c|^2
  \leq w (1 - w)$, then for a given value of $w$ the entropy is
  maximal for the state with $c = 0$ and for these states the entropy
  is equal to $S(\rho) = -w \ln w - (1 - w) \ln (1 - w)$.  When $|c|^2
  = w (1 - w)$ the states are pure and their entropy is $S(\rho) = 0$.
  
  The eigenvalues of $D_\Omega(\rho)$ are $w$ and $1 - w$, so the von
  Neumann entropy after the destruction is
  \begin{displaymath}
    S(D_\Omega(\rho)) = -\Tr\left(D_\Omega(\rho) \ln(D_\Omega(\rho))\right) 
    = -w \ln w - (1 - w) \ln (1 - w)
  \end{displaymath}
  and $S(D_\Omega(\rho)) \geq S(\rho)$, as it was expected from the
  theorem that the measurements with no selection increase entropy
  (see \cite{nielsen00}).
  
  Note that the destruction \emph{with selection} gives in this case
  \begin{displaymath}
    D^s_\Omega(\rho) = \cases{|0\rangle \langle0| 
      & if measurement of \(\Pi_\Omega\) gives \(1\)\\
      |\downarrow\rangle \langle\downarrow| 
      & if measurement of \(\Pi_\Omega\) gives \(0\).}
  \end{displaymath}
  Thus $S(D^s_\Omega(\rho)) = 0$ and we have
  \begin{displaymath}
    S(D^s_\Omega(\rho)) \leq S(\rho)
  \end{displaymath}
  i.e.\ the destruction with selection can decrease entropy.
\end{exa}

\section{Destruction in two-particle system---the space of states}
\label{sec:hilbert-space}

Now we discuss the space of states necessary for the description of
destruction of two-particle states of particles `$a$' and `$b$'.  Let
$\mathcal{H}_a$ and $\mathcal{H}_b$ be the Hilbert spaces for the
particle `$a$' and `$b$', respectively.  The two-particle Hilbert
space is the tensor product $\mathcal{H}_a \otimes \mathcal{H}_b$.
The state of the system is then described by the density matrix
$\rho$, which is an endomorphism of the space $\mathcal{H}_a \otimes
\mathcal{H}_b$, i.e.\ $\rho \in\End(\mathcal{H}_a \otimes
\mathcal{H}_b)$.  If one introduces in $\mathcal{H}_a$ an orthonormal
basis $\{|a\rangle\}$ and similarly in $\mathcal{H}_b$ an orthonormal
basis $\{|b\rangle\}$, then one can write the density matrix $\rho$ in
the form
\begin{equation}
  \label{eq:9}
  \rho = \sum_{aa'bb'} \rho_{aba'b'} (|a\rangle \otimes |b\rangle) 
  (\langle a'| \otimes \langle b'|) 
  = \sum_{aa'bb'} \rho_{aba'b'} |a\rangle \langle a'| 
  \otimes |b\rangle \langle b'|.
\end{equation}
In the case of identical particles the two-particle Hilbert space is,
of course, the projection onto the symmetric or antisymmetric part of
$\mathcal{H}_a \otimes \mathcal{H}_b$, thus we must additionally
require the appropriate behavior of the coefficients $\rho_{aba'b'}$
under the exchange of indices, i.e.\ 
\numparts%
\begin{eqnarray}
  \label{eq:10}\label{eq:10a}
  \rho_{aba'b'} = \rho_{baa'b'} = \rho_{abb'a'} = \rho_{bab'a'} \qquad
  &({\rm symmetric\; case})\\
  \label{eq:10b}
  \rho_{aba'b'} = -\rho_{baa'b'} = -\rho_{abb'a'} = \rho_{bab'a'} \qquad
  &({\rm antisymmetric\; case}).
\end{eqnarray}
\endnumparts

But such a description of composite quantum system is not enough if we
consider the measurement by the apparatus (mentioned in previous
sections) which can destroy the state.  The reason is that the density
matrix \eref{eq:9} can describe only the two-particle states of the
system, while after such a measurement we could have either a
one-particle state which evolves in time or a vacuum state.

This issue can be easily solved as in the case of one particle (see
\sref{sec:destr-one-part}), i.e.\ by introducing the one-dimensional
vacuum space $\mathcal{H}^0 \equiv \cov\{|0\rangle\}$, and taking the
direct sums $\mathcal{H}_a \oplus \mathcal{H}^0$ and $\mathcal{H}_b
\oplus \mathcal{H}^0$ instead of $\mathcal{H}_a$ and $\mathcal{H}_b$,
respectively.  The corresponding tensor product space can be
decomposed in the obvious way
\begin{eqnarray}
  \label{eq:11}
  \fl (\mathcal{H}_a \oplus \mathcal{H}^0) \otimes 
  (\mathcal{H}_b \oplus \mathcal{H}^0)
  \nonumber\\
  \lo= (\mathcal{H}_a \otimes \mathcal{H}_b) \oplus 
  \left((\mathcal{H}_a \otimes \mathcal{H}^0) 
    \oplus (\mathcal{H}^0 \otimes \mathcal{H}_b)\right) 
  \oplus (\mathcal{H}^0 \otimes \mathcal{H}^0). 
\end{eqnarray}
The first term on the right hand side of \eref{eq:11}, i.e.\ 
$\mathcal{H}_a \otimes \mathcal{H}_b$, describes two-particle states;
the second and third terms, i.e.\ $(\mathcal{H}_a \otimes
\mathcal{H}^0) \oplus (\mathcal{H}^0 \otimes \mathcal{H}_b)$,
represent one-particle states; while the last term, $\mathcal{H}^0
\otimes \mathcal{H}^0$, is the zero-particle state.  In the case of
distinguishable particles we can take the terms $\mathcal{H}_a \otimes
\mathcal{H}^0$ or $\mathcal{H}^0 \otimes \mathcal{H}_b$ as the Hilbert
space of the system after destruction of the particle `$b$' or `$a$',
respectively.  For identical particles we have to consider the
one-particle Hilbert space as a subspace of the sum $(\mathcal{H}
\otimes \mathcal{H}^0) \oplus (\mathcal{H}^0 \otimes \mathcal{H})$,
where $\mathcal{H}_a = \mathcal{H}_b \equiv \mathcal{H}$, because we
do not know if the particle `$a$' or `$b$' was destroyed.

The bases in the endomorphism spaces of the mentioned two-, one- and
zero-particle Hilbert spaces are
\numparts %
\begin{eqnarray}
  \label{eq:12}\label{eq:12a}
  (|a\rangle \otimes |b\rangle) (\langle a'| \otimes \langle b'|) 
  = |a\rangle \langle a'| \otimes |b\rangle \langle b'| &
  (\End(\mathcal{H}_a \otimes \mathcal{H}_b))\\
  \label{eq:12b}
  (|a\rangle \otimes |0\rangle) (\langle a'| \otimes \langle0|) 
  = |a\rangle \langle a'| \otimes |0\rangle \langle0| &
  (\End(\mathcal{H}_a \otimes \mathcal{H}^0))\\
  \label{eq:12c}
  (|0\rangle \otimes |b\rangle) (\langle0| \otimes \langle b'|) 
  = |0\rangle \langle0| \otimes |b\rangle \langle b'| &
  (\End(\mathcal{H}^0 \otimes \mathcal{H}_b))\\
  \label{eq:12d}
  (|0\rangle \otimes |0\rangle) (\langle0| \otimes \langle0|) 
  = |0\rangle \langle0| \otimes |0\rangle \langle0| \qquad&
  (\End(\mathcal{H}^0 \otimes \mathcal{H}^0)).
\end{eqnarray}
\endnumparts %
\emph{In the case of identical particles} $\mathcal{H}_a =
\mathcal{H}_b = \mathcal{H}$ and we consider \emph{the same basis} in
$\mathcal{H}_a$ and $\mathcal{H}_b$, i.e.\ $\{|a\rangle\} =
\{|b\rangle\}$.  The basis maps \eref{eq:12a}--\eref{eq:12d} should be
then supplemented by the basis endomorphisms
\numparts%
\begin{eqnarray}
  \label{eq:13}\label{eq:13a}
  (|a\rangle \otimes |0\rangle) (\langle0| \otimes \langle b'|) 
  = |a\rangle \langle0| \otimes |0\rangle \langle b'|\\
  \label{eq:13b}
  (|0\rangle \otimes |b\rangle) (\langle a'| \otimes \langle0|) 
  = |0\rangle \langle a'| \otimes |b\rangle \langle0|
\end{eqnarray}
\endnumparts %
which intertwine vectors from $\mathcal{H} \otimes \mathcal{H}^0$ to
$\mathcal{H}^0 \otimes \mathcal{H}$ and vice versa.

We point out that $\dim((\mathcal{H} \otimes \mathcal{H}^0) \oplus
(\mathcal{H}^0 \otimes \mathcal{H})) = 2 \dim(\mathcal{H} \otimes
\mathcal{H}^0)$, so for identical particles we must choose an
irreducible subspace of $(\mathcal{H} \otimes \mathcal{H}^0) \oplus
(\mathcal{H}^0 \otimes \mathcal{H})$ which corresponds to the space of
one-particle states.

\section{Supertraces}
\label{sec:operator-traces}

The partial traces $\Tr_a\colon \End(\mathcal{H}_a \otimes
\mathcal{H}_b) \to \End(\mathcal{H}_b)$ and $\Tr_b\colon
\End(\mathcal{H}_a \otimes \mathcal{H}_b) \to
\End(\mathcal{H}_a)$ are widely used in various contexts (see e.g.\ 
\cite{ballentine98}), but they cannot be used for the description of
the destruction.  Thus, our purpose is, in an analogy to
definition~\ref{df:op-trace}, to introduce maps that preserve the
trace and map $\End(\mathcal{H}_a \otimes \mathcal{H}_b)$ to
$\End(\mathcal{H}^0 \otimes \mathcal{H}^0)$, $\End(\mathcal{H}_a \otimes
\mathcal{H}^0)$ or $\End(\mathcal{H}^0 \otimes \mathcal{H}_b)$.

Let us start with the map $\End(\mathcal{H}_a \otimes \mathcal{H}_b) \to
\End(\mathcal{H}^0 \otimes \mathcal{H}^0)$.  Of course, we have
\begin{equation}
  \label{eq:16}
  \fl \End(\mathcal{H}_a \otimes \mathcal{H}_b) \ni 
  |\chi\rangle \langle\phi| \otimes |\psi\rangle \langle\xi| 
  \mapsto c |0\rangle \langle0| \otimes |0\rangle \langle0| \in 
  \End(\mathcal{H}^0 \otimes \mathcal{H}^0).
\end{equation}
The condition that the trace must be preserved leads to {$c =
  \Tr(|\chi\rangle \langle\phi| \otimes |\psi\rangle \langle\xi|) =
  \langle\phi|\chi\rangle \langle\xi|\psi\rangle$}, so we can define
the following linear map\footnote{We use the same symbol
  $\widehat{\Tr}$ for the map $\widehat{\Tr}\colon \End(\mathcal{H})
  \to \End(\mathcal{H}^0)$ and for the map $\widehat{\Tr}\colon
  \End(\mathcal{H}_a \otimes \mathcal{H}_b) \to
  \End(\mathcal{H}^0 \otimes \mathcal{H}^0)$, because the second map is
  the generalization of the first one in the tensor product space
  case.}:
\begin{df}
  \label{df:op-trace-tp}
  The tensor product supertrace $\widehat{\Tr}\colon
  \End(\mathcal{H}_a \otimes \mathcal{H}_b) \to \End(\mathcal{H}^0
  \otimes \mathcal{H}^0)$ is a linear map such that
  \begin{equation}
    \label{eq:17}
    \widehat{\Tr}(|\chi\rangle \langle\phi| 
    \otimes |\psi\rangle \langle\xi|) 
    = \langle\phi|\chi\rangle \langle\xi|\psi\rangle 
    (|0\rangle \langle0| \otimes |0\rangle \langle0|)
  \end{equation}
  for any $|\chi\rangle, |\phi\rangle \in \mathcal{H}_a$ and
  $|\psi\rangle, |\xi\rangle \in \mathcal{H}_b$.  Because of
  linearity, we can extend this map on the whole space
  $\End(\mathcal{H}_a \otimes \mathcal{H}_b)$.
\end{df}

Next, we need maps which transform the two-particle state into
one-particle state.  They are given by the following definition.
\begin{df}
  \label{df:partial-traces}
  The linear maps:
  \begin{description}
  \item[{\it left partial supertrace}] $\widehat{\Tr}_L\colon
    \End(\mathcal{H}_a \otimes \mathcal{H}_b) \to \End(\mathcal{H}^0
    \otimes \mathcal{H}_b)$,
  \item[{\it right partial supertrace}] $\widehat{\Tr}_R\colon
    \End(\mathcal{H}_a \otimes \mathcal{H}_b) \to \End(\mathcal{H}_a \otimes
    \mathcal{H}^0)$,
  \item[{\it inner partial supertrace}] $\widehat{\Tr}_I\colon
    \End(\mathcal{H}_a \otimes \mathcal{H}_b) \to 
    \End\left((\mathcal{H}_a \otimes
      \mathcal{H}^0) \oplus (\mathcal{H}^0 \otimes
      \mathcal{H}_b)\right)$,
  \item[{\it external partial supertrace}] $\widehat{\Tr}_E\colon
    \End(\mathcal{H}_a \otimes \mathcal{H}_b) \to 
    \End\left((\mathcal{H}_a \otimes
      \mathcal{H}^0) \oplus (\mathcal{H}^0 \otimes
      \mathcal{H}_b)\right)$,
  \end{description}
  act on the endomorphisms of the form $|\chi\rangle \langle\phi|
  \otimes |\psi\rangle \langle\xi| \in
  \End(\mathcal{H}_a \otimes \mathcal{H}_b)$ in the following way
  \numparts%
  \begin{eqnarray}
    \label{eq:22}
    \widehat{\Tr}_L(|\psi\rangle \langle\chi| 
    \otimes |\phi\rangle \langle\xi|) 
    = \langle\chi|\psi\rangle\, (|0\rangle \langle0| 
    \otimes |\phi\rangle \langle\xi|)\\
    \label{eq:23}
    \widehat{\Tr}_R(|\psi\rangle \langle\chi| 
    \otimes |\phi\rangle \langle\xi|) 
    = \langle\xi|\phi\rangle\, (|\psi\rangle \langle\chi| 
    \otimes |0\rangle \langle0|)\\
    \label{eq:24}
    \widehat{\Tr}_I(|\psi\rangle \langle\chi| 
    \otimes |\phi\rangle \langle\xi|) 
    = \langle\chi|\phi\rangle\, (|\psi\rangle \langle0| 
    \otimes |0\rangle \langle\xi|)\\
    \label{eq:25}
    \widehat{\Tr}_E(|\psi\rangle \langle\chi| 
    \otimes |\phi\rangle \langle\xi|) 
    = \langle\xi|\psi\rangle\, (|0\rangle \langle\chi| 
    \otimes |\phi\rangle \langle0|).    
  \end{eqnarray}
  \endnumparts%
  Because these superoperators are linear we can extend their action
  on the whole space $\End(\mathcal{H}_a \otimes \mathcal{H}_b)$ since
  every element of $\End(\mathcal{H}_a \otimes \mathcal{H}_b)$ can be
  written as the linear combination of the endomorphisms of the form
  $|\psi\rangle \langle\chi| \otimes |\phi\rangle \langle\xi|$.
\end{df}

We can see from \eref{eq:24} and \eref{eq:25} that the internal and
external partial supertraces $\widehat{\Tr}_I$ and $\widehat{\Tr}_E$
are non-trivial only for identical particles, i.e.\ for symmetric or
antisymmetric part of $\End\left((\mathcal{H} \otimes \mathcal{H}^0)
  \oplus (\mathcal{H}^0 \otimes \mathcal{H})\right)$ (notice that in
this case $\mathcal{H}_a = \mathcal{H}_b \equiv \mathcal{H}$), because
in the other case $\langle\chi|\phi\rangle$ and
$\langle\xi|\psi\rangle$ must vanish for any $|\psi\rangle,
|\chi\rangle \in \mathcal{H}_a$ and $|\phi\rangle, |\xi\rangle \in
\mathcal{H}_b$.

If we specify orthonormal bases $\{|a\rangle\}$ and $\{|b\rangle\}$ in
the spaces $\mathcal{H}_a$ and $\mathcal{H}_b$, respectively, then
\numparts%
\begin{eqnarray}
  \label{eq:22:1}
  \widehat{\Tr}_L\left(|a\rangle \langle a'| 
    \otimes |b\rangle \langle b'|\right) 
  = \delta_{a'a} |0\rangle \langle0| \otimes |b\rangle \langle b'|\\
  \label{eq:23:1}
  \widehat{\Tr}_R\left(|a\rangle \langle a'| 
    \otimes |b\rangle \langle b'|\right) 
  = \delta_{b'b} |a\rangle \langle a'| \otimes |0\rangle \langle0|\\
  \label{eq:24:1}
  \widehat{\Tr}_I\left(|a\rangle \langle a'| 
    \otimes |b\rangle \langle b'|\right) 
  = \delta_{a'b} |a\rangle \langle0| \otimes |0\rangle \langle b'|\\
  \label{eq:25:1}
  \widehat{\Tr}_E\left(|a\rangle \langle a'| 
    \otimes |b\rangle \langle b'|\right) 
  = \delta_{b'a} |0\rangle \langle a'| \otimes |b\rangle \langle0|.
\end{eqnarray}
\endnumparts

\begin{rem}
  \label{rem:traces}
  Let us note that the tensor product supertrace $\widehat{\Tr}$ from
  the definition~\ref{df:op-trace-tp} can be constructed as the
  following composition of partial supertraces
  \begin{eqnarray*}
    \widehat{\Tr} = \widehat{\Tr}_L \circ \widehat{\Tr}_R 
    = \widehat{\Tr}_R \circ \widehat{\Tr}_L\\
    \widehat{\Tr} = \widehat{\Tr}_I \circ \widehat{\Tr}_E 
    = \widehat{\Tr}_E \circ \widehat{\Tr}_I.
  \end{eqnarray*}
\end{rem}
\begin{rem}
  \label{rem:many-part}
  The definition~\ref{df:partial-traces} can be easily generalized to
  the case of states of more than two particles.  In such a case it is
  better to denote the partial supertraces by $\widehat{\Tr}_{ij}$,
  where we make the scalar product from $i$th vector (ket) and $j$th
  co-vector (bra) and replace them by $|0\rangle$ and $\langle0|$,
  respectively.  In such a notation we have $\widehat{\Tr}_R \equiv
  \widehat{\Tr}_{22}$, $\widehat{\Tr}_L \equiv \widehat{\Tr}_{11}$,
  $\widehat{\Tr}_I \equiv \widehat{\Tr}_{21}$, $\widehat{\Tr}_E \equiv
  \widehat{\Tr}_{12}$.  The partial supertraces which put more than
  one pair of $|0\rangle$ and $\langle0|$ can be easily obtained by
  taking an appropriate compositions of the partial supertraces
  $\widehat{\Tr}_{ij}$.
\end{rem}

\begin{lem}
  \label{lem:lr-pos-def}
  If $\sigma \in \End(\mathcal{H}_a \otimes \mathcal{H}_b)$ is
  non-negative then $\widehat{\Tr}_L(\sigma)$ and
  $\widehat{\Tr}_R(\sigma)$ are non-negative.
\end{lem}
\begin{proof}
  Let us show that $\widehat{\Tr}_L(\sigma)$ is non-negative for a
  non-negative $\sigma$.  Because $\widehat{\Tr}_L(\sigma) \in
  \End(\mathcal{H}^0 \otimes \mathcal{H}_b)$, we must show that 
  $(\langle0| \otimes \langle\phi|) \widehat{\Tr}_L(\sigma) (|0\rangle
  \otimes |\phi\rangle) \geq 0$ for any $|\phi\rangle \in
  \mathcal{H}_b$.  Without loss of generality we can assume that
  $|\phi\rangle$ is normalized, i.e.\ $\langle\phi|\phi\rangle = 1$.
  $\mathcal{H}_b$ can be decomposed into the linear covering of
  $|\phi\rangle$ and the subspace $\mathcal{H}_b^\perp$ of vectors
  orthogonal to $|\phi\rangle$.  If the set $\{|\tilde{b}\rangle\}$ is
  an orthonormal basis in $\mathcal{H}_b^\perp$, then the vector
  $|\phi\rangle$ and vectors from $\{|\tilde{b}\rangle\}$ make an
  orthonormal basis in $\mathcal{H}_b$.  Using $\sigma$ written in the
  basis $\{|a\rangle\}$ in $\mathcal{H}_a$ and the above basis in
  $\mathcal{H}_b$ and with help of \eref{eq:22:1} we get
  \begin{equation}
    \label{eq:29}
    (\langle0| \otimes \langle\phi|) \widehat{\Tr}_L(\sigma) 
    (|0\rangle \otimes |\phi\rangle) = \sum_{a} \sigma_{a\phi a\phi}
  \end{equation}
  where $\sigma_{a\phi a\phi} = (\langle a| \otimes \langle\phi|)
  \sigma (|a\rangle \otimes |\phi\rangle) \geq 0$ which follows from
  the assumption that $\sigma$ is non-negative.  Thus, indeed,
  non-negativeness of $\sigma$ implies non-negativeness of
  $\widehat{\Tr}_L(\sigma)$.  The proof for $\widehat{\Tr}_R(\sigma)$
  is analogous.
\end{proof}

Note that the analogous proof of non-negativeness for the usual
partial traces can be found e.g.\ in \cite{ballentine98}.

\section{Destruction in the system of two distinguishable particles}
\label{sec:destr-syst-two-d}

Now we consider the destruction of two-particle system of
distinguishable particles.  Let a density matrix of the form
\eref{eq:9} describes a system of two distinguishable particles `$a$'
and `$b$'.  The apparatus mentioned in \sref{sec:introduction}
destroys the particles if the outcomes of measurements of the
observables $\hat{\Lambda}_a$ and $\hat{\Lambda}_b$ lie in the subsets
$\Omega_a$ and $\Omega_b$ of spectra $\Lambda_a$ of $\hat{\Lambda}_a$
and $\Lambda_b$ of $\hat{\Lambda}_b$, respectively.  Let
$\Pi_{\Omega_a}$ be the projector onto the subspace of $\mathcal{H}_a$
associated with $\Omega_a$ and $\Pi_{\Omega_b}$ be the projector onto
the subspace of $\mathcal{H}_b$ associated with $\Omega_b$.  Now we
perform a simultaneous measurement of the observables $\Pi_{\Omega_a}
\otimes I_b$ and $I_a \otimes \Pi_{\Omega_b}$ ($I_a$ and $I_b$ denote
the identity operators in $\mathcal{H}_a$ and $\mathcal{H}_b$,
respectively).  Thus just after the measurement we have the following
four possible outcomes:
\begin{enumerate}
\item the measurement of $\Pi_{\Omega_a} \otimes I_b$ and $I_a \otimes
  \Pi_{\Omega_b}$ both give $0$---there are no particles to destroy
  and the final state is a two-particle state;
\item the measurement of $\Pi_{\Omega_a} \otimes I_b$ gives $0$ and
  the measurement of $I_a \otimes \Pi_{\Omega_b}$ gives $1$---the
  particle `$b$' is to destroy and the final state is a one-particle
  state of the particle `$a$';
\item the measurement of $\Pi_{\Omega_a} \otimes I_b$ gives $1$ and
  the measurement of $I_a \otimes \Pi_{\Omega_b}$ gives $0$---the
  particle `$a$' is to destroy and the final state is a one-particle
  state of the particle `$b$';
\item the measurement of $\Pi_{\Omega_a} \otimes I_b$ and $I_a \otimes
  \Pi_{\Omega_b}$ both give $1$---the particles `$a$' and `$b$' are to
  destroy and the final state is the vacuum state.
\end{enumerate}
One can easily verify the operators $\Pi_{\Omega_a}^\perp \otimes
\Pi_{\Omega_b}^\perp$, $\Pi_{\Omega_a}^\perp \otimes \Pi_{\Omega_b}$,
$\Pi_{\Omega_a} \otimes \Pi_{\Omega_b}^\perp$ and $\Pi_{\Omega_a}
\otimes \Pi_{\Omega_b}$, where $\Pi_{\Omega_a}^\perp \equiv I_a -
\Pi_{\Omega_a}$ and $\Pi_{\Omega_b}^\perp \equiv I_b - \Pi_{\Omega_b}$
are projectors on mutually orthogonal subspaces associated with the
cases (i)--(iv), respectively.  The probabilities for each of these
four situations are $\Tr\left[\rho (\Pi_{\Omega_a}^\perp \otimes
  \Pi_{\Omega_b}^\perp)\right]$, $\Tr\left[\rho (\Pi_{\Omega_a}^\perp
  \otimes \Pi_{\Omega_b})\right]$, $\Tr\left[\rho (\Pi_{\Omega_a}
  \otimes \Pi_{\Omega_b}^\perp)\right]$ and $\Tr\left[\rho
  (\Pi_{\Omega_a} \otimes \Pi_{\Omega_b})\right]$, respectively.

Now, in an analogy to the definitions~\ref{df:destr-sel}
and~\ref{df:dest1}, to destruct $\Omega_a$- and $\Omega_b$-projected
parts of the density matrix $\rho$ we apply appropriately the
$\widehat{\Tr}_L$ ($\widehat{\Tr}_R$) to the $\Omega_a$- ($\Omega_b$-)
projected part of $\rho$ as well as $\widehat{\Tr}$ to the $\Omega_a$-
and $\Omega_b$-projected part, and we arrive at the following
definitions.

\begin{df}
  \label{df:dest2n-sel}
  The destruction with selection in the set $\Omega$ of two-particle
  state $\rho \in \End(\mathcal{H}_a \otimes \mathcal{H}_b)$ of
  distinguishable particles is defined by the map $D^s_\Omega\colon
  \End(\mathcal{H}_a \otimes \mathcal{H}_b) \to \End(\mathcal{H}_a \otimes
  \mathcal{H}_b) \oplus \End(\mathcal{H}_a \otimes \mathcal{H}^0) \oplus
  \End(\mathcal{H}^0 \otimes \mathcal{H}_b) \oplus \End(\mathcal{H}^0 \otimes
  \mathcal{H}^0)$ of the form
  \begin{equation}
    \label{eq:37}
    D^s_\Omega(\rho) = 
    \cases{\frac{(\Pi_{\Omega_a}^\perp \otimes \Pi_{\Omega_b}^\perp) 
        \rho (\Pi_{\Omega_a}^\perp \otimes \Pi_{\Omega_b}^\perp)}%
      {\Tr[\rho (\Pi_{\Omega_a}^\perp \otimes \Pi_{\Omega_b}^\perp)]} 
      & for outcome (i)\\
      \frac{\widehat{\Tr}_R[(\Pi_{\Omega_a}^\perp \otimes \Pi_{\Omega_b}) 
        \rho (\Pi_{\Omega_a}^\perp \otimes \Pi_{\Omega_b})]}%
      {\Tr[\rho (\Pi_{\Omega_a}^\perp \otimes \Pi_{\Omega_b})]} 
      & for outcome (ii)\\ 
      \frac{\widehat{\Tr}_L[(\Pi_{\Omega_a} \otimes \Pi_{\Omega_b}^\perp) 
        \rho (\Pi_{\Omega_a} \otimes \Pi_{\Omega_b}^\perp)]}%
      {\Tr[\rho (\Pi_{\Omega_a} \otimes \Pi_{\Omega_b}^\perp)]} 
      & for outcome (iii)\\
      \frac{\widehat{\Tr}[(\Pi_{\Omega_a} \otimes \Pi_{\Omega_b}) 
        \rho (\Pi_{\Omega_a} \otimes \Pi_{\Omega_b})]}%
      {\Tr[\rho (\Pi_{\Omega_a} \otimes \Pi_{\Omega_b})]} 
      & for outcome (iv).}
  \end{equation}
\end{df}

\begin{df}
  \label{df:dest2n}
  The destruction with no selection in the set $\Omega$ of
  two-particle state $\rho \in \End(\mathcal{H}_a \otimes
  \mathcal{H}_b)$ of distinguishable particles is defined by the map
  $D_\Omega\colon
  \End(\mathcal{H}_a \otimes \mathcal{H}_b) \to \End(\mathcal{H}_a \otimes
  \mathcal{H}_b) \oplus
  \End(\mathcal{H}_a \otimes \mathcal{H}^0) \oplus \End(\mathcal{H}^0 \otimes
  \mathcal{H}_b) \oplus \End(\mathcal{H}^0 \otimes \mathcal{H}^0)$,
  such that
  \begin{eqnarray}
    \label{eq:016}
    \fl D_\Omega(\rho) = (\Pi_{\Omega_a}^\perp \otimes \Pi_{\Omega_b}^\perp) 
    \rho (\Pi_{\Omega_a}^\perp \otimes \Pi_{\Omega_b}^\perp) 
    + \widehat{\Tr}_R[(\Pi_{\Omega_a}^\perp \otimes \Pi_{\Omega_b}) 
    \rho (\Pi_{\Omega_a}^\perp \otimes \Pi_{\Omega_b})] 
    \nonumber\\
    + \widehat{\Tr}_L[(\Pi_{\Omega_a} \otimes \Pi_{\Omega_b}^\perp) 
    \rho (\Pi_{\Omega_a} \otimes \Pi_{\Omega_b}^\perp)]
    + \widehat{\Tr}[(\Pi_{\Omega_a} \otimes \Pi_{\Omega_b}) 
    \rho (\Pi_{\Omega_a} \otimes \Pi_{\Omega_b})]
  \end{eqnarray}
\end{df}

\begin{prop}
  \label{prop:dest2n}
  The superoperators $D^s_\Omega$ and $D_\Omega$ from the
  definitions~\textup{\ref{df:dest2n-sel}} and
  \textup{\ref{df:dest2n}}, respectively, are Kraus maps.
\end{prop}
\begin{proof}
  The verification that $D^s_\Omega(\rho)$ and $D_\Omega(\rho)$ are
  Hermitian is trivial.  Taking the density matrix $\rho$ in the form
  \eref{eq:9} one can easily check by straightforward calculation that
  $\Tr(D^s_\Omega(\rho)) = \Tr(\rho) = 1$ for every outcome (i)--(iv).
  Now,
  \begin{eqnarray}
    \label{eq:14}
    \fl \Tr\left(D_\Omega(\rho)\right) = \Tr[\rho (\Pi_{\Omega_a}^\perp 
    \otimes \Pi_{\Omega_b}^\perp)] 
    + \Tr[\rho (\Pi_{\Omega_a}^\perp \otimes \Pi_{\Omega_b})] 
    + \Tr[\rho (\Pi_{\Omega_a} \otimes \Pi_{\Omega_b}^\perp)] \nonumber\\ 
    + \Tr[\rho (\Pi_{\Omega_a} \otimes \Pi_{\Omega_b})] = \Tr(\rho) = 1.
  \end{eqnarray}
  $(\Pi_{\Omega_a}^\perp \otimes \Pi_{\Omega_b}^\perp) \rho
  (\Pi_{\Omega_a}^\perp \otimes \Pi_{\Omega_b}^\perp)$ is an
  orthogonal projection of a non-negative $\rho$, so it is
  non-negative.  Similarly, the entries $(\Pi_{\Omega_a}^\perp \otimes
  \Pi_{\Omega_b}) \rho (\Pi_{\Omega_a}^\perp \otimes \Pi_{\Omega_b})$
  and $(\Pi_{\Omega_a} \otimes \Pi_{\Omega_b}^\perp) \rho
  (\Pi_{\Omega_a} \otimes \Pi_{\Omega_b}^\perp)$ are non-negative.
  Therefore, using lemma~\ref{lem:lr-pos-def} we can see that
  $\widehat{\Tr}_R[(\Pi_{\Omega_a}^\perp \otimes \Pi_{\Omega_b}) \rho
  (\Pi_{\Omega_a}^\perp \otimes \Pi_{\Omega_b})]$ and
  $\widehat{\Tr}_L[(\Pi_{\Omega_a} \otimes \Pi_{\Omega_b}^\perp) \rho
  (\Pi_{\Omega_a} \otimes \Pi_{\Omega_b}^\perp)]$ are non-negative.
  $\widehat{\Tr}[(\Pi_{\Omega_a} \otimes \Pi_{\Omega_b}) \rho
  (\Pi_{\Omega_a} \otimes \Pi_{\Omega_b})]$ can be written as
  $\Tr[\rho (\Pi_{\Omega_a} \otimes \Pi_{\Omega_b})] |0\rangle
  \langle0| \otimes |0\rangle \langle0|$ and it is non-negative
  because $\Tr[\rho (\Pi_{\Omega_a} \otimes \Pi_{\Omega_b})] \geq 0$.
  Thus $D^s_\Omega(\rho)$ is non-negative.  Since all these four terms
  act in mutually orthogonal subspaces, $D_\Omega(\rho)$ is
  non-negative, too.  Therefore $D^s_\Omega$ and $D_\Omega$ are Kraus
  maps.
\end{proof}

Now, we illustrate the destruction of two-particle system of
distinguishable particles by the following example.
\begin{exa}
  Consider the system of spin-1 and spin-0 particles.  We assume that
  the destruction \emph{with no selection} takes place if the
  $z$-component of the spin of each particle is $0$.  We have
  $\hat{\Lambda}_a = \hat{S}_{a3}$ and $\hat{\Lambda}_b =
  \hat{S}_{b3}$.  So $\Lambda_a = \{-1, 0, 1\}$, $\Lambda_b = \{0\}$
  and $\Omega_a = \{0\}$, $\Omega_b = \{0\}$ (note that $\Omega_b =
  \Lambda_b$, so the outcomes (i) and (ii) are excluded).  We can take
  $\mathcal{H}_a = \cov\{|1,1\rangle, |1,0\rangle, |1,-1\rangle\}$ and
  $\mathcal{H}_b = \cov\{|0,0\rangle\}$, where $|j,m\rangle$ are the
  basis vectors.  The projectors can be written as $\Pi_{\Omega_a} =
  |1,0\rangle \langle1,0|$, $\Pi_{\Omega_a}^\perp = |1,1\rangle
  \langle1,1| + |1,-1\rangle \langle1,-1|$ and $\Pi_{\Omega_b} =
  |0,0\rangle \langle0,0|$.  The most general density matrix for such
  a state is
  \begin{eqnarray*}
    \fl \rho = w_1 |1,1\rangle \langle1,1| \otimes |0,0\rangle \langle0,0| 
    + c_1 |1,1\rangle \langle1,0| \otimes |0,0\rangle \langle0,0| 
    + c_2 |1,1\rangle \langle1,-1| \otimes |0,0\rangle \langle0,0|\\
    + c_1^* |1,0\rangle \langle1,1| \otimes |0,0\rangle \langle0,0| 
    + (1 - w_1 - w_2) |1,0\rangle \langle1,0| 
    \otimes |0,0\rangle \langle0,0|\\ 
    + c_3 |1,0\rangle \langle1,-1| \otimes |0,0\rangle \langle0,0| 
    + c_2^* |1,-1\rangle \langle1,1| \otimes |0,0\rangle \langle0,0|\\ 
    + c_3^* |1,-1\rangle \langle1,0| \otimes |0,0\rangle \langle0,0| 
    + w_2 |1,-1\rangle \langle1,-1| \otimes |0,0\rangle \langle0,0|
  \end{eqnarray*}
  where the coefficients $w_1, w_2 \in [0, 1]$, $c_1, c_2, c_3 \in
  \setC$ and they are restricted by the requirement that the density
  matrix $\rho$ is non-negative.  After the destruction we get a new
  state
  \begin{eqnarray*}
    \fl D_\Omega(\rho) = w_1 |1,1\rangle \langle1,1| 
    \otimes |0\rangle (0| 
    + w_2 |1,-1\rangle \langle1,-1| \otimes |0\rangle \langle0|\\ 
    + (1 - w_1 - w_2) |0\rangle \langle0| \otimes |0\rangle \langle0|
  \end{eqnarray*}
  (recall that $|0\rangle$ denotes the \emph{vacuum} vector), so the
  new state is a mixture of the spin-1 particle in up direction (with
  the probability $w_1$), the spin-1 particle in down direction (with
  the probability $w_2$) and the vacuum (with the probability $1 - w_1
  - w_2$).
\end{exa}

\section{Destruction in the system of two identical particles}
\label{sec:destr-syst-two-i}

Now we consider the destruction in the state of two identical
particles.  In this case $\mathcal{H}_a = \mathcal{H}_b \equiv
\mathcal{H}$.  The system of two identical particles is described by a
density matrix of the form \eref{eq:9} together with the symmetry
conditions \eref{eq:10a} or \eref{eq:10b}.  As in the previous cases,
let $\Pi_\Omega$ be the projector onto the subspace of $\mathcal{H}$
associated with $\Omega \subset \Lambda$.  Now we perform a
measurement of the symmetrized observable $\Pi_\Omega \otimes I + I
\otimes \Pi_\Omega$.  The spectral decomposition of this observable is
\begin{equation}
  \label{eq:2}
  \fl \Pi_\Omega \otimes I + I \otimes \Pi_\Omega 
  = 0 \cdot \Pi_\Omega^\perp \otimes \Pi_\Omega^\perp 
  + 1 \cdot (\Pi_\Omega^\perp \otimes \Pi_\Omega 
  + \Pi_\Omega \otimes \Pi_\Omega^\perp) 
  + 2 \cdot \Pi_\Omega \otimes \Pi_\Omega
\end{equation}
($\Pi_\Omega^\perp = I - \Pi_\Omega$, as before), where
\begin{description}
\item[$\Pi_\Omega^\perp \otimes \Pi_\Omega^\perp$] corresponds to the
  situation that there is no particle with an eigenvalue of
  $\hat{\Lambda}$ belonging to $\Omega$,
\item[$\Pi_\Omega^\perp \otimes \Pi_\Omega + \Pi_\Omega \otimes
  \Pi_\Omega^\perp$] corresponds to the situation that there is
  exactly one particle with an eigenvalue of $\hat{\Lambda}$ belonging
  to $\Omega$,
\item[$\Pi_\Omega \otimes \Pi_\Omega$] corresponds to the situation
  that there are two particles with an eigenvalue of $\hat{\Lambda}$
  belonging to $\Omega$.
\end{description}
In view of \eref{eq:2}, just after the measurement, we have only the
three possibilities:
\begin{enumerate}
\item the measurement of $\Pi_\Omega \otimes I + I \otimes \Pi_\Omega$
  gives $0$---there is no particle to destroy and the final state is a
  two-particle state,
\item the measurement of $\Pi_\Omega \otimes I + I \otimes \Pi_\Omega$
  gives $1$---there is exactly one particle to destroy and the final
  state is a one-particle state,
\item the measurement of $\Pi_\Omega \otimes I + I \otimes \Pi_\Omega$
  gives $2$---there are two particle to destroy and the final state is
  the vacuum state.
\end{enumerate}
The probabilities that one of the three cases (i)--(iii) occurs are
$\Tr[\rho (\Pi_\Omega^\perp \otimes \Pi_\Omega^\perp)]$, $\Tr[\rho
(\Pi_\Omega^\perp \otimes \Pi_\Omega + \Pi_\Omega \otimes
\Pi_\Omega^\perp)]$ and $\Tr[\rho (\Pi_\Omega \otimes \Pi_\Omega)]$,
respectively.

In order to destruct the $\Omega$-projected part of the density matrix
$\rho$ we apply the same algorithm as in the case of distinguishable
particles, but now we cannot omit $\widehat{\Tr}_I$ and
$\widehat{\Tr}_E$ because their action is non-trivial.  Therefore, we
can formulate the following definitions.
\begin{df}
  \label{df:dest2i-sel}
  The destruction with selection in the set $\Omega$ of two-particle
  state $\rho \in \End(\mathcal{H} \otimes \mathcal{H})$ of identical
  particles is defined by the map $D^s_\Omega\colon \End(\mathcal{H}
  \otimes \mathcal{H}) \to
  \End(\mathcal{H} \otimes \mathcal{H}) \oplus \End((\mathcal{H}
  \otimes \mathcal{H}^0) \oplus (\mathcal{H}^0 \otimes \mathcal{H}))
  \oplus
  \End(\mathcal{H}^0 \otimes \mathcal{H}^0)$, such that
  \numparts%
  \begin{eqnarray}
    \label{eq:7}
    \fl D^s_\Omega(\rho) = \frac{(\Pi_\Omega^\perp \otimes \Pi_\Omega^\perp) 
      \rho (\Pi_\Omega^\perp \otimes \Pi_\Omega^\perp)}%
    {\Tr[\rho (\Pi_\Omega^\perp \otimes \Pi_\Omega^\perp)]} \\
    \label{eq:15}
    \fl D^s_\Omega(\rho) = \left\{\widehat{\Tr}_R[(\Pi_\Omega^\perp 
      \otimes \Pi_\Omega) \rho (\Pi_\Omega^\perp \otimes \Pi_\Omega)]
      + \widehat{\Tr}_L[(\Pi_\Omega \otimes \Pi_\Omega^\perp) 
      \rho (\Pi_\Omega \otimes \Pi_\Omega^\perp)]\right. \nonumber\\
    \left. \pm \widehat{\Tr}_I[(\Pi_\Omega^\perp \otimes \Pi_\Omega) 
      \rho (\Pi_\Omega \otimes \Pi_\Omega^\perp)]
      \pm \widehat{\Tr}_E[(\Pi_\Omega \otimes \Pi_\Omega^\perp) 
      \rho (\Pi_\Omega^\perp \otimes \Pi_\Omega)]\right\} \nonumber\\
     \times \left\{\Tr[\rho (\Pi_\Omega^\perp \otimes \Pi_\Omega)] 
       + \Tr[\rho (\Pi_\Omega \otimes \Pi_\Omega^\perp)]\right\}^{-1}\\
     \label{eq:19}
     \fl D^s_\Omega(\rho) = \frac{\widehat{\Tr}[(\Pi_\Omega 
       \otimes \Pi_\Omega) \rho (\Pi_\Omega \otimes \Pi_\Omega)]}%
     {\Tr[ \rho (\Pi_\Omega \otimes \Pi_\Omega)]}
  \end{eqnarray}
  \endnumparts%
  for the outcomes (i), (ii) and (iii) of the measurement of
  $\Pi_\Omega \otimes I + I \otimes \Pi_\Omega$, respectively; where
  the signs $+$ and $-$ correspond to symmetric and antisymmetric
  cases, respectively.
\end{df}
\begin{df}
  \label{df:dest2i}
  The destruction with no selection in the set $\Omega$ of
  two-particle state $\rho \in
  \End(\mathcal{H} \otimes \mathcal{H})$ of identical particles is 
  defined by the map $D_\Omega\colon \End(\mathcal{H} \otimes
  \mathcal{H}) \to \End(\mathcal{H} \otimes \mathcal{H}) \oplus
  \End((\mathcal{H} \otimes \mathcal{H}^0) \oplus (\mathcal{H}^0
  \otimes \mathcal{H})) \oplus
  \End(\mathcal{H}^0 \otimes \mathcal{H}^0)$, such that
  \begin{eqnarray}
    \label{eq:35}
    \fl D_\Omega(\rho) = (\Pi_\Omega^\perp \otimes \Pi_\Omega^\perp) 
    \rho (\Pi_\Omega^\perp \otimes \Pi_\Omega^\perp)
    + \widehat{\Tr}_R[(\Pi_\Omega^\perp \otimes \Pi_\Omega) 
    \rho (\Pi_\Omega^\perp \otimes \Pi_\Omega)] 
    \nonumber\\
    + \widehat{\Tr}_L[(\Pi_\Omega \otimes \Pi_\Omega^\perp) 
    \rho (\Pi_\Omega \otimes \Pi_\Omega^\perp)]
    \pm \widehat{\Tr}_I[(\Pi_\Omega^\perp \otimes \Pi_\Omega) 
    \rho (\Pi_\Omega \otimes \Pi_\Omega^\perp)]
    \nonumber\\
    \pm \widehat{\Tr}_E[(\Pi_\Omega \otimes \Pi_\Omega^\perp) 
    \rho (\Pi_\Omega^\perp \otimes \Pi_\Omega)]
    + \widehat{\Tr}[(\Pi_\Omega \otimes \Pi_\Omega) 
    \rho (\Pi_\Omega \otimes \Pi_\Omega)]
  \end{eqnarray}
  where the signs $+$ and $-$ correspond to symmetric and
  antisymmetric cases, respectively.
\end{df}

In view of the discussion at the end of \sref{sec:hilbert-space}, we
shall show the following lemma.
\begin{lem}
  \label{lem:irred}
  For a symmetric or antisymmetric density matrix $\rho \in
  \End(\mathcal{H} \otimes \mathcal{H})$ the 
  state given by \textup{\eref{eq:15}} belongs to the
  \emph{irreducible} one-particle subspace of $\End((\mathcal{H}
  \otimes \mathcal{H}^0) \oplus (\mathcal{H}^0 \otimes \mathcal{H}))$
  \textup(the signs $+$ and $-$ correspond to symmetric and
  antisymmetric cases, respectively\textup).
\end{lem}
\begin{proof}
  Let the sets of vectors $\{|\beta\rangle\}$ and $\{|\alpha\rangle\}$
  be the orthonormal basis in $\mathcal{H}_\Omega$ and
  $\mathcal{H}_\Omega^\perp$, respectively.  So the set
  $\{|\alpha\rangle\} \cup \{|\beta\rangle\}$ is a basis in
  $\mathcal{H}$.  Let us write the density matrix $\rho$ in the
  form~\eref{eq:9} using this basis.  Moreover, we can write
  $\Pi_\Omega = \sum_\beta |\beta\rangle \langle\beta|$ and
  $\Pi_\Omega^\perp = \sum_\alpha |\alpha\rangle \langle\alpha|$.
  Therefore, using the symmetry conditions \eref{eq:10a} or
  \eref{eq:10b}, we get
  \begin{eqnarray}
    \label{eq:30}
    \fl \widehat{\Tr}_R[(\Pi_\Omega^\perp \otimes \Pi_\Omega) 
    \rho (\Pi_\Omega^\perp \otimes \Pi_\Omega)] 
    + \widehat{\Tr}_L[(\Pi_\Omega \otimes \Pi_\Omega^\perp) 
    \rho (\Pi_\Omega \otimes \Pi_\Omega^\perp)]
    \nonumber\\
    \pm \widehat{\Tr}_I[(\Pi_\Omega^\perp \otimes \Pi_\Omega) 
    \rho (\Pi_\Omega \otimes \Pi_\Omega^\perp)]
    \pm \widehat{\Tr}_E[(\Pi_\Omega \otimes \Pi_\Omega^\perp) 
    \rho (\Pi_\Omega^\perp \otimes \Pi_\Omega)]
    \nonumber\\
    \lo= \sum_{\alpha\alpha'} \left(\sum_\beta 
      \rho_{\alpha\beta\alpha'\beta}\right) 
    \left(|\alpha\rangle \otimes |0\rangle 
      + |0\rangle \otimes |\alpha\rangle\right) 
    \left(\langle\alpha'| \otimes \langle0| 
      + \langle0| \otimes \langle\alpha'|\right)
  \end{eqnarray}
  so, it belongs to one-particle irreducible subspace of
  $\End((\mathcal{H} \otimes \mathcal{H}^0) \oplus (\mathcal{H}^0
  \otimes \mathcal{H}))$.
\end{proof}

\begin{prop}
  \label{prop:dest2i}
  The superoperators $D^s_\Omega$ and $D_\Omega$ from the
  definitions~\textup{\ref{df:dest2i-sel}} and
  \textup{\ref{df:dest2i}}, respectively, are Kraus maps.
\end{prop}
\begin{proof}
  To prove that $D^s_\Omega(\rho)$ and $D_\Omega(\rho)$ are Hermitian,
  we have only to check if the sum $\widehat{\Tr}_I[(\Pi_\Omega^\perp
  \otimes \Pi_\Omega) \rho (\Pi_\Omega \otimes \Pi_\Omega^\perp)] +
  \widehat{\Tr}_E[(\Pi_\Omega \otimes \Pi_\Omega^\perp) \rho
  (\Pi_\Omega^\perp \otimes \Pi_\Omega)]$ is Hermitian, since the
  remaining parts of \eref{eq:7} or \eref{eq:35} are evidently
  Hermitian.  First, observe that $\left((\Pi_\Omega^\perp \otimes
    \Pi_\Omega) \rho (\Pi_\Omega \otimes
    \Pi_\Omega^\perp)\right)^{\dag} = (\Pi_\Omega \otimes
  \Pi_\Omega^\perp) \rho (\Pi_\Omega^\perp \otimes \Pi_\Omega)$.  Now,
  it is easy to see from the definition~\ref{df:partial-traces} that
  for any endomorphism $\sigma \in \End(\mathcal{H} \otimes
  \mathcal{H})$ we have $\left(\widehat{\Tr}_I(\sigma)\right)^{\dag} =
  \widehat{\Tr}_E(\sigma^{\dag})$ and \emph{vice versa}.  Therefore
  \begin{eqnarray}
    \label{eq:8}
    \fl \left(\widehat{\Tr}_I[(\Pi_\Omega^\perp \otimes \Pi_\Omega) 
      \rho (\Pi_\Omega \otimes \Pi_\Omega^\perp)] +
      \widehat{\Tr}_E[(\Pi_\Omega \otimes \Pi_\Omega^\perp) 
      \rho (\Pi_\Omega^\perp \otimes \Pi_\Omega)]\right)^{\dag} 
    \nonumber\\
    \lo= \widehat{\Tr}_E[(\Pi_\Omega \otimes \Pi_\Omega^\perp) 
    \rho (\Pi_\Omega^\perp \otimes \Pi_\Omega)] 
    + \widehat{\Tr}_I[(\Pi_\Omega^\perp \otimes \Pi_\Omega) 
    \rho (\Pi_\Omega \otimes \Pi_\Omega^\perp)].
  \end{eqnarray}
  Thus $D^s_\Omega(\rho)$ and $D_\Omega(\rho)$ are Hermitian.  In
  order to prove that $\Tr\left(D^s_\Omega(\rho)\right) =
  \Tr\left(D_\Omega(\rho)\right) = \Tr(\rho)$ it is enough to notice
  that the diagonal elements of the internal and external partial
  supertraces vanish.  This is evident from \eref{eq:24:1} and
  \eref{eq:25:1}.  In virtue of this fact, the rest of the proof of
  this point is analogous to the proof of the respective part of
  proposition~\ref{prop:dest2n}.  $(\Pi_\Omega^\perp \otimes
  \Pi_\Omega^\perp) \rho (\Pi_\Omega^\perp \otimes \Pi_\Omega^\perp)$
  and $\widehat{\Tr}[(\Pi_\Omega \otimes \Pi_\Omega) \rho (\Pi_\Omega
  \otimes \Pi_\Omega)]$ are of course non-negative.  The proof that
  the sum
  \begin{eqnarray}
    \label{eq:1}
    \fl \widehat{\Tr}_R[(\Pi_\Omega^\perp \otimes \Pi_\Omega) 
    \rho (\Pi_\Omega^\perp \otimes \Pi_\Omega)] 
    + \widehat{\Tr}_L[(\Pi_\Omega \otimes \Pi_\Omega^\perp) 
    \rho (\Pi_\Omega \otimes \Pi_\Omega^\perp)]
    \nonumber\\ 
    \pm \widehat{\Tr}_I[(\Pi_\Omega^\perp \otimes \Pi_\Omega) 
    \rho (\Pi_\Omega \otimes \Pi_\Omega^\perp)] 
    \pm \widehat{\Tr}_E[(\Pi_\Omega \otimes \Pi_\Omega^\perp) 
    \rho (\Pi_\Omega^\perp \otimes \Pi_\Omega)]
  \end{eqnarray}
  is non-negative is the following.  Let $|\phi\rangle \otimes
  |0\rangle + |0\rangle \otimes |\phi\rangle$ be the vector from
  $(\mathcal{H} \otimes \mathcal{H}^0) \oplus (\mathcal{H}^0 \otimes
  \mathcal{H})$.  The vector $|\phi\rangle \in \mathcal{H}$ can be
  decomposed as follows $|\phi\rangle = c |x\rangle + d |y\rangle$,
  where $|x\rangle \in \mathcal{H}_\Omega^\perp$, $|y\rangle \in
  \mathcal{H}_\Omega$, $c, d \in \setC$ and $\langle x|x\rangle =
  \langle y|y\rangle = 1$.  Next, we construct the basis in the
  subspace $\mathcal{H}_\Omega^\perp$ as in the proof of the
  lemma~\ref{lem:lr-pos-def}, with the vector $|x\rangle$ basis
  vector.  Now, using \eref{eq:30} we get
  \begin{eqnarray}
    \label{eq:31}
    \fl \left(\langle\phi| \otimes \langle0| 
      + \langle0| \otimes \langle\phi|\right) 
    \left(\widehat{\Tr}_R[(\Pi_\Omega^\perp \otimes \Pi_\Omega) 
      \rho (\Pi_\Omega^\perp \otimes \Pi_\Omega)] 
    + \widehat{\Tr}_L[(\Pi_\Omega \otimes \Pi_\Omega^\perp) 
    \rho (\Pi_\Omega \otimes \Pi_\Omega^\perp)]\right.
    \nonumber\\
    \fl \left. \pm \widehat{\Tr}_I[(\Pi_\Omega^\perp \otimes \Pi_\Omega) 
      \rho (\Pi_\Omega \otimes \Pi_\Omega^\perp)]
    \pm \widehat{\Tr}_E[(\Pi_\Omega \otimes \Pi_\Omega^\perp) 
    \rho (\Pi_\Omega^\perp \otimes \Pi_\Omega)]\right) 
  \left(|\phi\rangle \otimes |0\rangle + |0\rangle \otimes |\phi\rangle\right)
  \nonumber\\
  \lo= 4 |c|^2 \sum_\beta \rho_{x\beta x\beta}.
  \end{eqnarray}
  Clearly the sum in \eref{eq:31} is non-negative, since $\rho$ is
  non-negative.  Thus $D^s_\Omega(\rho)$ is also non-negative.  Since
  the sum \eref{eq:1} and the other terms in \eref{eq:35} act in
  mutually orthogonal subspaces, $D_\Omega(\rho)$ is also
  non-negative.  Therefore $D^s_\Omega$ and $D_\Omega$ are Kraus maps.
\end{proof}

Now, we illustrate the destruction of two-particle system of identical
particles by the following example.
\begin{exa}
  Consider the system of two identical spin-$\frac{1}{2}$ particles.
  We assume that the destruction \emph{with no selection} takes place
  if the $z$-component of the spin of any particle is $\frac{1}{2}$.
  The observable $\hat{\Lambda}$, its spectrum $\Lambda$, the subset
  $\Omega$, one-particle Hilbert space $\mathcal{H}$ as well as
  projectors $\Pi_\Omega$ and $\Pi_\Omega^\perp$ are the same as in
  example~\ref{exa:destr-1}.  The two-particle space of states is
  antisymmetric part of $\mathcal{H} \otimes \mathcal{H}$, i.e.\ 
  $\cov\{\frac{1}{\sqrt{2}} (|\uparrow\rangle \otimes
  |\downarrow\rangle -\allowbreak |\downarrow\rangle \otimes
  |\uparrow\rangle)\}$.  This space is one-dimensional, thus the state
  is a pure one, and its density matrix is of the form
  \begin{displaymath}
    \fl \rho = \frac{1}{2} \left(|\uparrow\rangle \langle\uparrow| 
      \otimes |\downarrow\rangle \langle\downarrow| 
      - |\uparrow\rangle \langle\downarrow| 
      \otimes |\downarrow\rangle \langle\uparrow| 
      - |\downarrow\rangle \langle\uparrow| 
      \otimes |\uparrow\rangle \langle\downarrow| 
      + |\downarrow\rangle \langle\downarrow| 
      \otimes |\uparrow\rangle \langle\uparrow|\right).
  \end{displaymath}
  After the destruction of the particles with $S_3 = \frac{1}{2}$, we
  get the new state
  \begin{eqnarray*}
    \fl D_\Omega(\rho) 
    = \frac{1}{2} \left(|\downarrow\rangle \langle\downarrow| 
      \otimes |0\rangle \langle0| 
      + |0\rangle \langle0| 
      \otimes |\downarrow\rangle \langle\downarrow| 
      + |\downarrow\rangle \langle0| 
      \otimes |0\rangle \langle\downarrow| 
      + |0\rangle \langle\downarrow| 
      \otimes |\downarrow\rangle \langle0|\right)\\
    \lo= \frac{1}{2} \left(|\downarrow\rangle  \otimes |0\rangle 
      + |0\rangle \otimes |\downarrow\rangle\right) 
    \left(\langle\downarrow|  \otimes \langle0| 
      + \langle0| \otimes \langle\downarrow|\right).
  \end{eqnarray*}
  So it is really an element of one-dimensional irreducible subspace
  of $\End((\mathcal{H} \otimes \mathcal{H}^0) \oplus (\mathcal{H}^0
  \otimes \mathcal{H}))$.
  
  It should be noted that in this case the destruction \emph{with
    selection} gives the same result.
  
  Because before and after the destruction we deal with pure states
  the von Neumann entropies of the initial and destroyed states are
  both equal zero.
\end{exa}

\section{Conclusions}
\label{sec:conclusions}

We have given a mathematical formalism which allows one to describe
the destruction of a particle from the two-particle state in the
framework of quantum mechanics.  This is done by means of the
reduction procedure \cite{neumann31} (with selection or with no
selection) associated with immediate mapping of the part of the
reduced density matrix onto vacuum density matrix and is based on the
use of supertraces.  We point out that the destruction procedure can
be treated as a supplement to the von Neumann--L{\"u}ders measurement
procedure.

Moreover, our formalism of destructions, developed for the case of
one-particle and two-particle states, can be uniquely generalized to
the multi-particle states, with help of the partial supertraces
$\widehat{\Tr}_{ij}$ (see remark~\ref{rem:many-part}).  Also, it can
be easily extended to the generalized measurements by means of
positive operator-valued measures (POVM) rather than orthogonal
projections.

The formalism introduced herein should be helpful in a description of
the processes when one has the system under time evolution after the
destruction.  This may happen in the Einstein--Podolsky--Rosen type
experiments (the destruction can take place in a detector).  For this
reason the destruction procedure may also be helpful in quantum
information theory.  The study of different destruction processes as
well as applications of the destruction procedure to calculation of
the EPR quantum correlations will be done in the forthcoming papers.

\ack

The authors would acknowledge useful comments from B Broda, T
Brzezi{\'n}ski, J K{\l}osi{\'n}ski and K Kowalski.  This work is
supported by the University of {\L}{\'o}d{\'z} grant.

\section*{References}

\end{document}